\documentclass[a4paper,11pt]{article}
\pdfoutput=1 

\usepackage{jheppub,amsmath,graphicx,hyperref} 

\usepackage[T1]{fontenc} 

\title{\boldmath Singular Instantons in Eddington-inspired-Born-Infeld Gravity}
\author[a]{Frederico Arroja,}
\author[a,b]{Che-Yu Chen,}
\author[a,b,c]{Pisin Chen}
\author[a]{and Dong-han Yeom}
\affiliation[a]{Leung Center for Cosmology and Particle Astrophysics, National Taiwan University, Taipei, Taiwan 10617}
\affiliation[b]{Department of Physics, National Taiwan University, Taipei, Taiwan 10617}
\affiliation[c]{Kavli Institute for Particle Astrophysics and Cosmology, SLAC National Accelerator Laboratory, Stanford University, Stanford, CA 94305, U.S.A.}
\emailAdd{arroja@phys.ntu.edu.tw}
\emailAdd{b97202056@gmail.com}
\emailAdd{pisinchen@phys.ntu.edu.tw}
\emailAdd{innocent.yeom@gmail.com}

\abstract{In this work, we investigate $O(4)$-symmetric instantons within the Eddington-inspired-Born-Infeld gravity theory (EiBI). We discuss the regular Hawking-Moss instanton and find that the tunneling rate reduces to the General Relativity (GR) value, even though the action value is different by a constant. We give a thorough analysis of the singular Vilenkin instanton and the Hawking-Turok instanton with a quadratic scalar field potential in the EiBI theory. In both cases, we find that the singularity can be avoided in the sense that the physical metric, its scalar curvature and the scalar field are regular under some parameter restrictions, but there is a curvature singularity of the auxiliary metric compatible with the connection. We find that the on-shell action is finite and the probability does not reduce to its GR value. We also find that the Vilenkin instanton in the EiBI theory would still cause the instability of the Minkowski space, similar to that in GR, and this is observationally inconsistent. This result suggests that the singularity of the auxiliary metric may be problematic at the quantum level and that these instantons should be excluded from the path integral.}

\begin{document}
\maketitle
\flushbottom

\section{Introduction\label{sec:intro}}
Even though Einstein's General Relativity (GR) is an extremely successful theory of gravity \cite{Misner:1974qy}, it has some fundamental problems in the sense that the theory inevitably contains singularities \cite{Penrose:1964wq}. At the singularity the theory breaks down and the equations of motion are no longer valid. From this perspective, it can be conjectured that GR, as a purely classical theory, to some extent is not a complete theory to describe our universe as a whole and some sort of quantum effects should come into play. However, a complete and self-consistent quantum theory of gravity is still an open question. In order to uncover the nature of quantum gravity, a possible phenomenological approach is to find a modified theory of gravity which can not only recover GR in the low curvature limit, but also naturally resolve the aforementioned singularity problem in GR. This modified theory of gravity can be seen as an effective theory of a more fundamental quantum theory of gravity, below some curvature cutoff scale. 

Following this direction, a modified theory of gravity dubbed Eddington-inspired-Born-Infeld theory was proposed in Ref.~\cite{Banados:2010ix}. This theory is equivalent to Einstein's GR in vacuum but differs from it when matter fields are included. Essentially, the motivation of this theory traces back to the huge success of the Born-Infeld action for classical electrodynamics in solving the divergence of the self-energy of point-like charges in 1934 \cite{Born:1934gh}. Then, in the pioneering work of Deser and Gibbons \cite{Deser:1998rj}, the Born-Infeld-inspired action was firstly applied to a gravitational setup. However, the theory proposed in \cite{Deser:1998rj} is constructed on a pure metric formalism, and it contains fourth order derivatives in the field equations and ghost instabilities. The EiBI theory, on the other hand, is constructed upon the Palatini formalism and it is free from the ghost problems. Since its proposal, the EiBI theory has been widely studied from astrophysical and cosmological points of view. In Refs.~\cite{Scargill:2012kg,Cho:2012vg,Avelino:2012ue} it was shown that the initial Big Bang singularity of the universe can be avoided. However, some classical singular states were found and this fact may indicate that the EiBI theory is also not completely singularity-free \cite{Pani:2012qd,Bouhmadi-Lopez:2013lha,Bouhmadi-Lopez:2014jfa,Bouhmadi-Lopez:2014tna}. Recently, it was shown that some of the singularities in this theory (e.g. big rip singularity) can be avoided by quantum effects \cite{Bouhmadi-Lopez:2016dcf}. Furthermore, the nonlinear coupling between gravity and matter in the EiBI theory was studied in Refs.~\cite{Delsate:2012ky,Cho:2013usa}. The properties of compact objects were investigated, including collapsing dust \cite{Pani:2011mg,Pani:2012qb}, neutron stars \cite{Harko:2013wka,Sham:2013cya,Sotani:2015tya}, black hole solutions \cite{Wei:2014dka,Sotani:2014lua,Sotani:2015ewa,Olmo:2013gqa,Jana:2015cha,Bazeia:2016rlg} and wormhole solutions \cite{Harko:2013aya,Shaikh:2015oha,Tamang:2015tmd,Olmo:2015bya,Olmo:2015dba}. Cosmological perturbations \cite{EscamillaRivera:2012vz,Yang:2013hsa} and large scale structure formation \cite{Du:2014jka} were also studied. An inflationary solution driven by an inflaton minimally coupled in the EiBI gravity was obtained and analyzed in Refs.~\cite{Cho:2013pea,Cho:2014xaa,Cho:2014jta,Cho:2015yua}. Some constraints on the parameter $\kappa$, which characterizes the theory, were obtained from the sun \cite{Casanellas:2011kf}, neutron stars \cite{Avelino:2012ge}, and nuclear physics \cite{Avelino:2012qe}. The tightest constraint obtained with atomic nuclei in Ref.~\cite{Avelino:2012qe} is $\kappa\lesssim 10^{-3} \mathrm{kg}^{-1} \mathrm{m}^{5} \mathrm{s}^{-2}$ (this constraint is comparable to the one obtained in Ref.~\cite{Avelino:2012ge} using neutron stars). Besides, in Ref.~\cite{Avelino:2012ge} it was found that the existence of an imaginary effective sound speed for negative $\kappa$ may lead to unwanted instabilities. Finally, in Refs.~\cite{Makarenko:2014lxa,Makarenko:2014nca,Fernandes:2014bka,Odintsov:2014yaa,Jimenez:2014fla,Makarenko:2014cca,Chen:2015eha,Elizalde:2016vsd,Bambi:2016xme} some extensions of the EiBI theory were proposed.

One major benefit of the EiBI theory is that it can avoid or resolve the initial singularity of the universe. In the EiBI theory, two effective metrics are introduced, where one is the physical metric that is non-dynamical but couples to Standard Model particles, while the other is the auxiliary metric that does not couple to matter fields but is dynamical. Due to this property, the singularity is in the dynamical part, the physical metric is regular and hence classical physical observers cannot see any effects from the singularity. In this sense, the EiBI theory can resolve the initial singularity problem \footnote{In fact, in Refs.~\cite{Bouhmadi-Lopez:2013lha,Bouhmadi-Lopez:2014jfa} the authors pointed out that even though the physical metric is regular at the beginning of the universe, the scalar curvature of the auxiliary metric diverges and the universe described by the auxiliary metric is singular. On the other hand, they also found that for some singularities at large scale, the auxiliary metric can be more regular than the physical metric.}.

In the classical sense, the resolution of the initial singularity via the separation of the physical metric and the auxiliary metric would be successful. However, it is fair to say that some sectors of the theory are still singular. Then the natural question to ask is: is the curvature singularity of the auxiliary metric acceptable in the quantum mechanical sense?

In order to deal with this problem, we have to investigate the theory of quantum gravity. Of course, we have no consensus on the correct theory of quantum gravity yet, but the \textit{Euclidean path-integral approach} \cite{EQG} would be a conservative and non-perturbative approach toward this. In this paper, we follow this approach to study the quantum gravitational properties of the EiBI theory. In particular, we will focus on \textit{singular instantons} in the EiBI theory.

In Einstein's gravity, the roles of the singular instantons are controversial. The Hawking-Turok instanton \cite{Hawking:1998bn} is a good example. The Hawking-Turok instanton is an instanton solution of a massive scalar field that allows an open universe. Usually, an open universe can be created via the Coleman-DeLuccia instanton \cite{Coleman:1980aw}. This process requires two or more vacua but the Hawking-Turok instanton does not require the existence of multiple vacua. The cost to pay is that the Hawking-Turok instanton is singular at the end point. Even though some curvature invariants are divergent at the end point, the action integral remains regular and hence a finite and well-defined probability for the Hawking-Turok instanton is attainable.

However, the allowance of such a singular instanton would cause some unacceptable problems, or at least, it needs more justifications. One very interesting criticism was put forward by Vilenkin \cite{Vilenkin:1998pp}. Vilenkin introduced a singular instanton of a free scalar field in the Minkowski background. Similar to the Hawking-Turok instanton, the Vilenkin instanton gives a well-defined probability. It implies that the Minkowski space is \textit{unstable} via such a singular instanton, whereas we believe that the Minkowski space should be stable if there is only a free scalar field. Therefore, the allowance of the singular instanton would be problematic.

A natural question to ask is: can the EiBI theory resolve the singularity problem of these instantons? In order to deal with this problem, we investigate the Vilenkin instanton as well as the Hawking-Turok instanton in the EiBI theory. If, similar to the initial singularity of the universe, one can resolve the singularity of the instantons (with some parameter restrictions) in the sense that the scalar curvature of the physical metric is regular despite the fact that that of the auxiliary metric is singular, then we may be able to say whether the Vilenkin and Hawking-Turok instantons in the EiBI framework are acceptable in the path integral or not.

This paper is outlined as follows. In section~\ref{sec:eibiaction} we give a concise introduction to the EiBI gravity theory, including its action, field equations, and an alternative action which is dynamically equivalent to the original action. In section~\ref{sec:eofandHM} we obtain the general equations of motion for instantons in the EiBI theory. A simple application to the Hawking-Moss tunneling instanton is presented as well. In section~\ref{secvilenkin} and section~\ref{secHT} we give a thorough analysis on the Vilenkin instanton and the Hawking-Turok instanton in the EiBI theory, respectively. We firstly review the singular behavior of these instantons in GR. In the EiBI setup we obtain the complete solutions and their on-shell actions numerically. Furthermore, the asymptotic solutions near the end points are presented analytically. We finally present our conclusions in section~\ref{sec:conclusion}.

\section{The EiBI model and the alternative action}
\label{sec:eibiaction}
We will study instanton solutions in the framework of the EiBI theory of gravity, whose action reads \cite{Banados:2010ix}
\begin{equation}
S_{EiBI}=\frac{1}{\kappa}\int d^4x\Big[\sqrt{|g_{\mu\nu}+\kappa R_{\mu\nu}(\Gamma)|}-\lambda\sqrt{-g}\Big]+S_M(g).
\label{actioneibi}
\end{equation}
The theory is constructed upon a Palatini formalism in which the metric $g_{\mu\nu}$ and the connection $\Gamma$ should be treated as independent variables. The connection is assumed to be torsionless and $R_{\mu\nu}(\Gamma)$ is chosen to be the symmetric part of the Ricci tensor, which is constructed solely by the independent connection. The expression under the square root in the first term is the absolute value of the determinant of the tensor $g_{\mu\nu}+\kappa R_{\mu\nu}(\Gamma)$. Furthermore, $g$ denotes the determinant of $g_{\mu\nu}$ and $S_M$ represents the matter Lagrangian, where matter is assumed to be coupled covariantly to $g_{\mu\nu}$ only. The Palatini formulation ensures that the field equations of the theory have derivatives up to only second order, even though the action of the theory can have a rather different form from that of GR. In addition, $\lambda$ is a dimensionless constant which relates to an effective cosmological constant of the theory at low curvature limit through $\Lambda\equiv(\lambda-1)/\kappa$. The parameter $\kappa$ is a constant characterizing the theory. In the limit of $\kappa\rightarrow 0$, the theory approaches GR. Note that we will work with reduced Planck units $8\pi G=1$ and set the speed of light to $c=1$.

Even though the action \eqref{actioneibi} is widely regarded as the starting action in the literature, the square root of the curvature in the action usually leads to technical challenges when canonical quantization of the theory is under consideration \cite{Bouhmadi-Lopez:2016dcf}. Therefore, in this work we will consider an alternative action, which is dynamically equivalent to action \eqref{actioneibi}. This alternative action is \cite{Delsate:2012ky}
\begin{equation}
S'_{EiBI}=\frac{1}{2}\int d^4x\sqrt{-q}\Big[R[q]-\frac{2}{\kappa}+\frac{1}{\kappa}\Big(q^{\alpha\beta}g_{\alpha\beta}-2\sqrt{\frac{g}{q}}\lambda\Big)\Big]+S_M(g),
\label{actionalternative}
\end{equation}
where $R[q]\equiv q^{\alpha\beta}R_{\beta\alpha}(q)$ is the scalar curvature of the auxiliary metric. In this alternative formulation, it is the auxiliary metric $q_{\mu\nu}$ and the physical metric $g_{\mu\nu}$ that are treated independently. The Ricci tensor $R_{\mu\nu}(q)$ is constructed from the Levi-Civita connection of the auxiliary metric $q_{\mu\nu}$. Note that $q^{\mu\nu}$ and $q$ are the inverse and the determinant of $q_{\mu\nu}$, respectively. This action is similar to a bi-gravity action without dynamics for $g_{\mu\nu}$ \cite{Delsate:2012ky}, and is dynamically equivalent to \eqref{actioneibi} in the sense that both actions lead to the same classical field equations. More precisely, the variations of action \eqref{actionalternative} with respect to $q_{\mu\nu}$ and $g_{\mu\nu}$ result in
\begin{subequations}\label{eq12}
\begin{align}
q_{\mu\nu}&=g_{\mu\nu}+\kappa R_{\mu\nu},\label{eq1}\\
\sqrt{-q}q^{\mu\nu}&=\lambda\sqrt{-g}g^{\mu\nu}-\sqrt{-g}\kappa T^{\mu\nu},\label{eq2}
\end{align}
\end{subequations}
respectively. In the above equations $T_{\mu\nu}$ stands for the energy-momentum tensor. Note that these equations of motion can be obtained by varying the action \eqref{actioneibi} with respect to the independent connection and the physical metric as well.
It is also interesting to note the resemblance of the previous equations with a disformal transformation.
The advantage of applying the action \eqref{actionalternative} is its resemblance to the Einstein-Hilbert action in the sense that it is linear on the auxiliary Ricci scalar $R[q]$. Therefore, this similarity enables us to obtain some important results in the EiBI theory, such as the boundary term in the Euclidean action, by directly analogizing to the corresponding results in GR. We will elucidate this in detail in the following sections.

\section{The Euclidean equations of motion and the on-shell action}\label{sec:eofandHM}
In order to investigate the instantons within the EiBI theory, in this section we present the Euclidean equations of motion by applying the alternative action \eqref{actionalternative}. Considering the EiBI gravity coupled to a scalar field $\phi$ with a potential $V(\phi)$, the Euclidean version of the action is
\begin{eqnarray}
S_E&=&-\frac{1}{2}\int d^4x\sqrt{+q}\Big[R[q]-\frac{2}{\kappa}+\frac{1}{\kappa}\Big(q^{\alpha\beta}g_{\alpha\beta}-2\sqrt{\frac{g}{q}}\lambda\Big)\Big]\nonumber\\
&&+\int d^4x\sqrt{+g}\Big(\frac{(\nabla\phi)^2}{2}+V(\phi)\Big)+S_{B}.
\label{euaction}
\end{eqnarray}
The last term $S_B$ is the boundary action, which is introduced to render a well-defined variation procedure. Because the alternative action is linear on $R[q]$ just like in GR, and due to the fact that the boundary terms in varying the action result only from the second order derivatives in $R[q]$, we can simply write down $S_B$ as the Gibbons-Hawking boundary action of the auxiliary metric $q_{\mu\nu}$ \cite{Gibbons:1976ue}:
\begin{equation}
S_B=-\int_{\partial M}d^3x\sqrt{+h(q)}K(q).
\label{gibbonboundaryq}
\end{equation}
Note that the $3D$ induced metric $h_{ab}(q)$ and the extrinsic curvature $K(q)$ on the boundary $\partial M$ are constructed solely by the auxiliary metric $q_{\mu\nu}$.

We consider $O(4)$-symmetric instantons which in general have dominating contributions in the path integral formulation of quantum gravity. These instantons can be described by two Euclidean minisuperspace metrics:
\begin{equation}
\begin{split}
ds_g^2=N(\tau)^2d\tau^2+a(\tau)^2d\Omega_3^2\,,
\qquad
ds_q^2=M(\tau)^2d\tau^2+b(\tau)^2d\Omega_3^2\,,
\end{split}
\end{equation}
where $d\Omega_3^2$ is the metric on a three-sphere. In the above equations, $a$ and $b$ ($N$ and $M$) are the scale factors (lapse functions) of the physical and auxiliary metrics, respectively. Then, the Ricci scalar of the auxiliary metric $q$ is
\begin{equation}
R[q]=6\Big(\frac{1}{b^2}-\frac{\dot{b}^2}{M^2b^2}-\frac{\ddot{b}}{M^2b}+\frac{\dot{b}}{b}\frac{\dot{M}}{M^3}\Big),
\end{equation}
where the dot denotes the derivative with respect to $\tau$. The minisuperspace action \eqref{euaction} becomes
\begin{eqnarray}
S_E&=&-2\pi^2\Big(\frac{1}{2}\Big)\int Mb^3d\tau\Big[6\Big(\frac{1}{b^2}+\frac{\dot{b}^2}{M^2b^2}\Big)-\frac{2}{\kappa}+\frac{1}{\kappa}\Big(\frac{N^2}{M^2}+3\frac{a^2}{b^2}-2\lambda\frac{Na^3}{Mb^3}\Big)\Big]\nonumber\\
&&+2\pi^2\int Na^3d\tau\Big(\frac{\dot{\phi}^2}{2N^2}+V(\phi)\Big)+\textrm{boundary terms},
\label{maction}
\end{eqnarray}
where an integration by part is used to obtain action \eqref{maction}. The factor $2\pi^2$ is the volume of the three-sphere. Note that the boundary terms in the last line of \eqref{maction} do not contribute to the results of variations because of the presence of $S_B$. Variations of the action \eqref{maction} with respect to $N$, $a$ and $M$ lead to the following constraint equations 
\begin{eqnarray}
\frac{N}{M}&=&\frac{a^3}{b^3}\Big[\lambda-\kappa\Big(\frac{\dot{\phi}^2}{2N^2}-V(\phi)\Big)\Big],\label{ee1}\\
\frac{M}{N}&=&\frac{a}{b}\Big[\lambda+\kappa\Big(\frac{\dot{\phi}^2}{2N^2}+V(\phi)\Big)\Big],\label{ee3}\\
\frac{\dot{b}^2}{M^2}&=&1-\frac{b^2}{3\kappa}\Big(1+\frac{N^2}{2M^2}-\frac{3a^2}{2b^2}\Big),\label{ee2}
\end{eqnarray}
respectively. Furthermore, variations of the action \eqref{maction} with respect to $b$ and $\phi$ result in the equations of motion
\begin{eqnarray}
0&=&\ddot{\phi}+\Big(3\frac{\dot{a}}{a}-\frac{\dot{N}}{N}\Big)\dot{\phi}-N^2\frac{dV}{d\phi},\label{eibiphi}\\
\ddot{b}&=&\frac{\dot{M}}{M}\dot{b}-\frac{M^2b}{3\kappa}+\frac{bN^2}{3\kappa},\label{equationb}
\end{eqnarray}
respectively. In general, we have proven that the equations of motion, Eqs.~\eqref{ee1}-\eqref{equationb}, are consistent. To be more precise, we can always fix the lapse function $N$ at will since one of these five equations is redundant. 

For the sake of later convenience, we rewrite Eqs.~\eqref{ee1} and \eqref{ee3} as
\begin{eqnarray}
\frac{b^4}{a^4}&=&\Big[(\lambda+\kappa V(\phi))^2-\frac{\kappa^2\dot{\phi}^4}{4N^4}\Big],\nonumber\\
\frac{M^4}{N^4}&=&\frac{\Big(\lambda+\kappa V(\phi)+\frac{\kappa\dot{\phi}^2}{2N^2}\Big)^4}{\Big[(\lambda+\kappa V(\phi))^2-\frac{\kappa^2\dot{\phi}^4}{4N^4}\Big]}.
\label{3.11}
\end{eqnarray}
The on-shell action can be obtained by inserting the equations of motion above into \eqref{euaction} to find:
\begin{eqnarray}
S_E&=&\int \frac{d^4x}{\kappa}\Big[\sqrt{+q}\Big(\frac{q^{\alpha\beta}g_{\alpha\beta}}{2}-1\Big)-\sqrt{+g}(\lambda+\kappa V(\phi))\Big]+S_B\nonumber\\
&=&\frac{2\pi^2}{\kappa}\int d\tau Mb^3\Big(\frac{a^2}{b^2}-1\Big)+S_B\nonumber\\
&=&\frac{2\pi^2}{\kappa}\int d\tau Na^3\Big(\lambda+\kappa V(\phi)+\frac{\kappa\dot{\phi}^2}{2N^2}\Big)\Big(1-\frac{b^2}{a^2}\Big)+S_B.
\end{eqnarray}
If we set $N=1$, the on-shell action becomes
\begin{equation}
S_E=\frac{2\pi^2}{\kappa}\int d\tau a^3\Big(\lambda+\kappa V(\phi)+\frac{\kappa\dot{\phi}^2}{2}\Big)\Big(1-\frac{b^2}{a^2}\Big)+S_B.
\label{bulkaction}
\end{equation}
Note that $S_B$ is the Gibbons-Hawking boundary term defined by $q$, i.e, Eq.~\eqref{gibbonboundaryq}.

\subsection{A simple example: Hawking-Moss instanton}
The equations of motion \eqref{ee1}-\eqref{equationb} are the most general equations which can be applied to all $O(4)$-symmetric instantons in the EiBI setup. For example, we can regard the scalar field $\phi$ as a tunneling field that undergoes a transition from different vacuum states on a potential $V(\phi)$. In this case, the potential can be assumed to have two local minima. One of them is the false vacuum $\phi_F$ and the other corresponds to the true vacuum $\phi_T$. Between these two local minima, there is a local maximum $\phi_{max}$. 

Among these tunneling instantons, in this subsection we will analyze the Hawking-Moss instanton \cite{Hawking:1981fz} to understand the tunneling process in the EiBI theory. In this case, the tunneling field is fixed at a stationary point $\phi_{s}$ on the potential where $\dot{\phi}=0$. Therefore, Eq.~\eqref{eibiphi} is trivially satisfied and the constraints \eqref{3.11} become
\begin{equation}
\frac{M^2}{N^2}=\frac{b^2}{a^2}=\lambda+\kappa V(\phi_{s})\equiv X^2,
\end{equation}
where $X$ is a non-zero constant. The constraint equation \eqref{ee2} can be rewritten as
\begin{equation}
\frac{\dot{b}^2}{M^2}=1-\frac{b^2}{3\kappa}\Big(1-\frac{1}{X^2}\Big)\equiv1-\frac{b^2}{3}\Lambda_q,
\end{equation}
where
\begin{equation}
\Lambda_q\equiv\frac{1}{\kappa}\Big(1-\frac{1}{X^2}\Big).
\end{equation}
The subscript $q$ is used to emphasize that the equation is constructed by the auxiliary metric $q_{\mu\nu}$. By assuming $M$ to be a constant, the solution of $b$ can be derived as follows
\begin{equation}
b=\sqrt{\frac{3}{\Lambda_q}}\sin{\left(M\sqrt{\frac{\Lambda_q}{3}}\tau\right)}.
\end{equation}
Finally, the on-shell action can be obtained by integrating from $\tau=0$ to $\tau=\pi\sqrt{3/\Lambda_q}/M$:
\begin{equation}
S_E=-\frac{24\pi^2}{\Lambda_q}.
\end{equation}
So we have
\begin{eqnarray}
S=-S_E=\frac{8\pi^2}{H_q^2}&=&\frac{A_q}{4l_P^2}\nonumber\\
&=&24\pi^2\kappa+\frac{A_g}{4l_P^2},
\label{hmentropy}
\end{eqnarray}
where $A_q=4\pi/H_q^2$ and $A_g=4\pi/H^2$ are the area of the cosmological horizon of the auxiliary metric and physical metric, respectively, and $S$ is the entropy of the instanton. Note that $H_q^2=\Lambda_q/3$ and $l_P$ is the Planck length which we introduce explicitly in the previous expressions so that they can be easily compared with the well-known results. From the second line of Eq.~\eqref{hmentropy}, one can see that the on-shell action value is different from its GR counterpart by a constant and this constant is linear on $\kappa$. The action value in the EiBI theory reduces exactly to that of GR as $\kappa\rightarrow0$. It is interesting to remark that in the EiBI theory, there are two ways to define the area of the horizon: either with the physical metric or with the auxiliary metric (the two expressions given in Eq.~\eqref{hmentropy}). Therefore, it is less clear whether the entropy of the Hawking-Moss instanton in the EiBI theory is proportional to the horizon area or not. However, the first law of thermodynamics would remain intact, because it requires only the variation of the entropy.

Furthermore, the tunneling rate, which is defined by $\textrm{exp}(-\Delta S_E)$, is equivalent to that in GR. Note that $\Delta S_E$ is the difference between the action of the instanton solution that satisfies the Euclidean equations of motion and that calculated at the location of the false vacuum. This can be seen from the fact that the extra constant term proportional to $\kappa$ cancels in $\Delta S_E$. This result is expected because the physical process and its relevant physical quantities (the tunneling probability in this case) in the EiBI theory should reduce to GR in vacuum, i.e., in the absence of a dynamical scalar field.

\section{Vilenkin singular instanton}\label{secvilenkin}
As mentioned in the introduction, the most interesting feature of the EiBI theory is its ability to cure or alleviate singularities in GR. Even though the tunneling rate for the Hawking-Moss instanton in the EiBI theory is the same as that in GR, it is nevertheless interesting to investigate the instanton solutions where the presence of a singularity is inevitable in GR. In those cases one can question whether the avoidance of singularities in the EiBI theory can still hold in the instanton setup. And if so, whether these regular instantons are acceptable in the EiBI theory? In this section we will analyze the Vilenkin instanton, the simplest singular instanton, proposed in Ref.~\cite{Vilenkin:1998pp} to test whether the singular states can be alleviated in the EiBI theory. We will also calculate the on-shell action of this instanton to see whether the instability problem of the Minkowski space can be resolved \cite{Vilenkin:1998pp}.

\subsection{GR case}\label{subsec:grvil}
Let us briefly review the Vilenkin instanton in GR. In Ref.~\cite{Vilenkin:1998pp}, the author considered a massless scalar field interacting with the gravitational field. The Euclidean action is
\begin{equation}
S_{E(GR)}=\int\sqrt{+g}d^4x\Big(-\frac{R(g)}{2}+\frac{(\nabla\phi)^2}{2}\Big)+S_{boundary},
\end{equation}
where $S_{boundary}$ corresponds to the Gibbons-Hawking action of $g_{\mu\nu}$ on the boundary. The Euclidean spacetime is
\begin{equation}
ds_g^2=d\tau^2+a^2d\Omega_3^2.
\end{equation}
Note that the lapse function $N$ is assumed to $N=1$. The field equations of $\phi$ and $a$, and the constraint equation are
\begin{eqnarray}
0&=&\ddot{\phi}+3\frac{\dot{a}}{a}\dot{\phi},\label{phieq}\\
\ddot{a}&=&-\frac{a}{3}\dot{\phi}^2,\label{bbbb}\\
\dot{a}^2&=&1+\frac{1}{6}a^2\dot{\phi}^2\label{constraint}.
\end{eqnarray}
In the Vilenkin instanton \cite{Vilenkin:1998pp}, the boundary conditions corresponding to an asymptotically flat instanton are
\begin{equation}
\begin{split}
a(\tau)\approx\tau\,,
\qquad
\phi(\tau)\rightarrow0\,,
\end{split}
\label{vilenboundarycondition}
\end{equation}
when $\tau\rightarrow\infty$. From Eq.~\eqref{phieq}, $\dot{\phi}$ can be solved
\begin{equation}
\dot{\phi}=-\frac{C}{a^3},
\label{phidot}
\end{equation}
where $C$ is an integration constant. Substituting it into Eq.~\eqref{constraint}, we have
\begin{equation}
\dot{a}^2=1+\frac{C^2}{6a^4}.\label{eqagr}
\end{equation}
The exact solutions of $\phi(\tau)$ and $a(\tau)$ were found in Ref.~\cite{GonzalezDiaz:1998du}. There is a curvature singularity at $\tau=\tau_s$ where the asymptotic behaviors of $a$ and $\phi$ can be written as
\begin{equation}
\begin{split}
\bar a^3(\bar\tau)\approx\sqrt{\frac{3}{2}}(\bar\tau-\bar\tau_s)\,,
\qquad
\phi(\bar\tau)\approx-\sqrt{\frac{2}{3}}\log{(\bar\tau-\bar\tau_s)}+\textrm{constant}\,.
\end{split}
\label{grvilenkinasy}
\end{equation}
In the above equations $\bar{a}$ and $\bar\tau$ are dimensionless quantities defined by $\bar{a}\equiv a/\sqrt{C}$ and $\bar\tau\equiv\tau/\sqrt{C}$.

In Ref.~\cite{Vilenkin:1998pp}, it was found that the contribution to the Euclidean action emanates from the Gibbons-Hawking boundary term:
\begin{equation}
S_{boundary}=\sqrt{6}\pi^2 C,
\end{equation}
which can be arbitrarily small by choosing a small $C$. Vilenkin therefore concluded that flat space is unstable because the nucleation probability is proportional to $\textrm{exp}(-S_{E(GR)})$, which is unsuppressed for a small $C$. Furthermore, he argued that this instability problem may jeopardize the validity of the singular Hawking-Turok instanton proposed in Ref.~\cite{Hawking:1998bn} because the Vilenkin-type singularity has a similar behavior to that of the Hawking-Turok type in GR. It is then interesting to see whether the aforementioned problems can be solved in the EiBI theory.

\subsection{EiBI case}
\begin{figure}[t]
\centering
\graphicspath{{fig/}}
\includegraphics[scale=0.8]{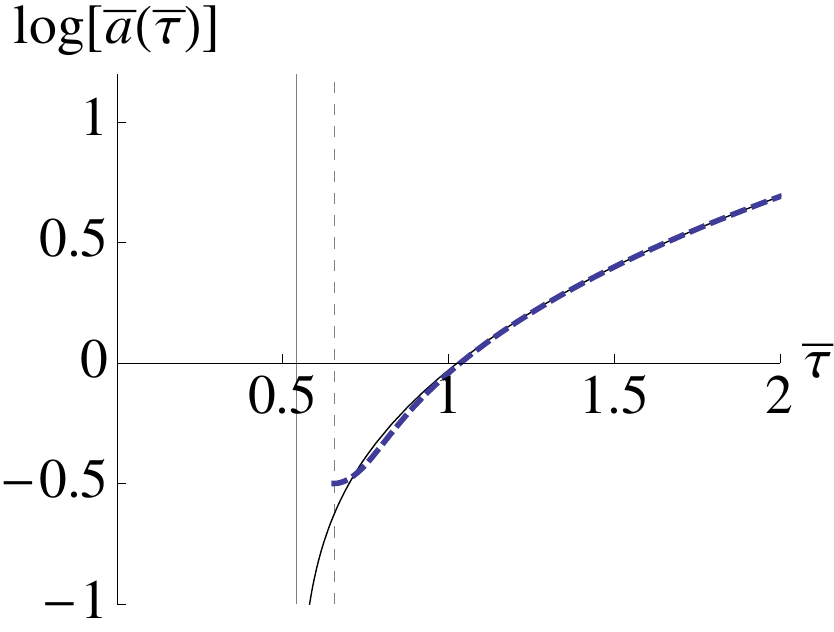}
\includegraphics[scale=0.8]{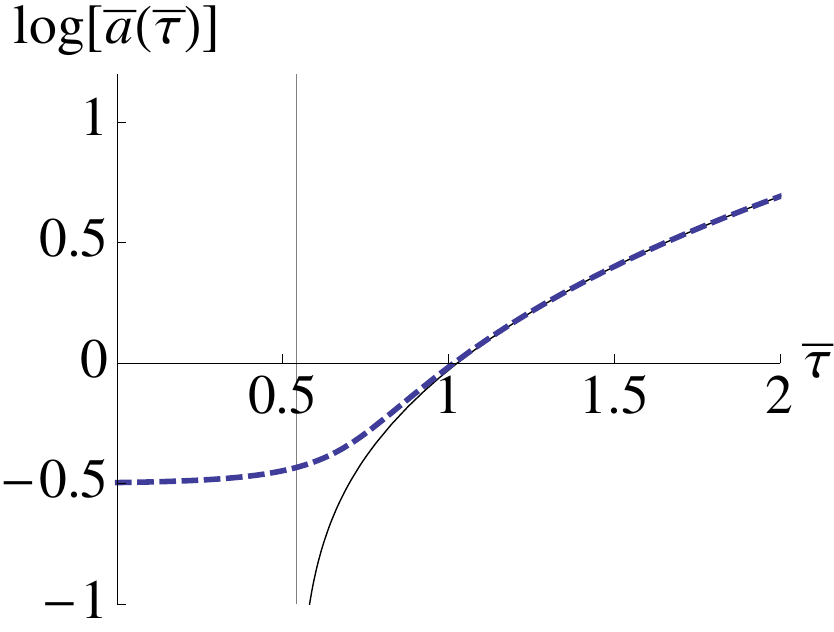}
\caption{The numerical results of $\log{\bar{a}(\bar\tau)}$ are shown in the dashed curves when $\bar\kappa$ is equal to $0.1$ (left) and $-0.1$ (right). The black solid curve is the singular Vilenkin's solution in GR and the vertical solid line indicates the singular point $\bar\tau_s$. If $\bar\kappa>0$, the scale factor reaches $\bar{a}_0$ at a finite $\bar{\tau}$ which is highlighted by the dashed vertical line. If $\bar\kappa<0$, the scale factor approaches $\bar{a}_0$ when $\bar{\tau}\rightarrow-\infty$. See Eqs.~\eqref{vilenapk} and \eqref{vilenanka}.}
\label{vilena}
\end{figure}

\begin{figure}[t]
\graphicspath{{fig/}}
\includegraphics[scale=0.8]{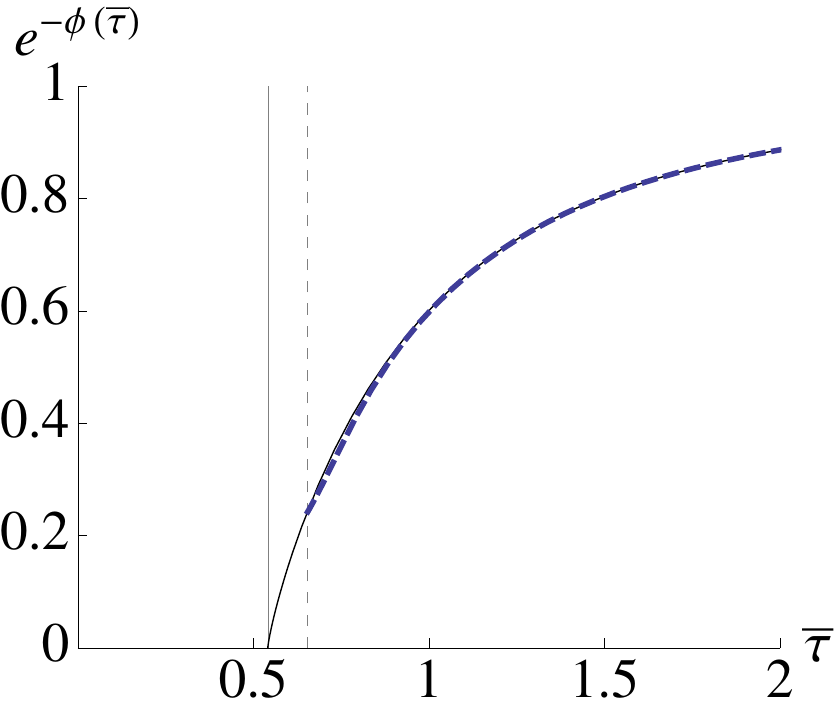}
\includegraphics[scale=0.8]{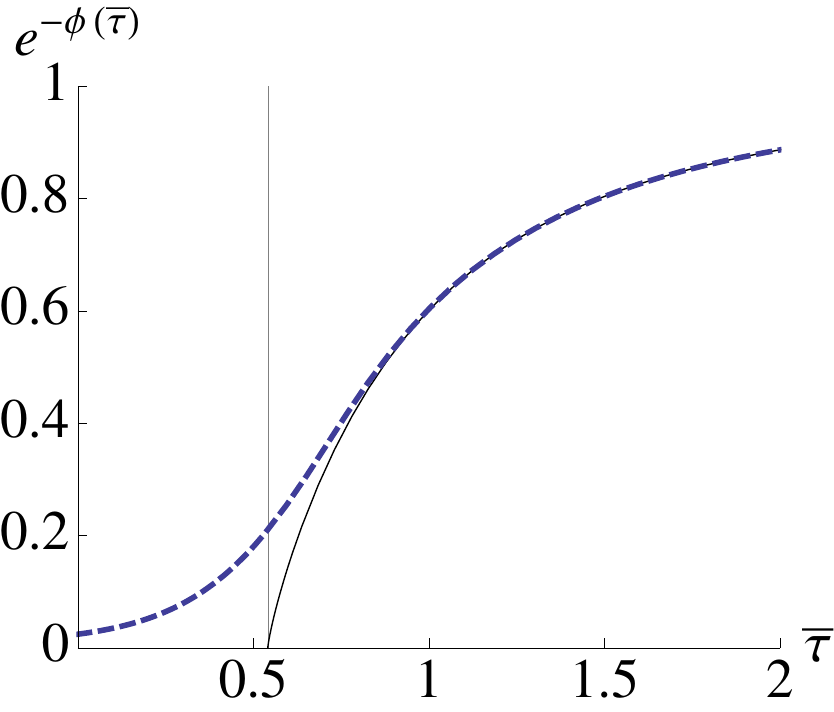}
\caption{The numerical results of $\textrm{e}^{-\phi(\bar\tau)}$ are shown in the dashed curves when $\bar\kappa$ is equal to $0.1$ (left) and $-0.1$ (right). The black solid curve is the singular Vilenkin's solution in GR. If $\bar\kappa>0$, the scalar field terminates with a non-zero slope. See Eq.~\eqref{phivilenpkkk}. If $\bar\kappa<0$, the asymptotic behavior of the scalar field is $\phi\propto-\bar\tau$. See Eq.~\eqref{phinkappa}.}
\label{vilenphi}
\end{figure}

In the Vilenkin instanton setup, the scalar field potential is assumed to be zero and the equation of motion of $\phi$, i.e., Eq.~\eqref{eibiphi}, reduces to Eq.~\eqref{phieq} when $N=1$. Therefore, Eq.~\eqref{phidot} is still valid in the EiBI theory and Eqs.~\eqref{3.11} can be rewritten as
\begin{eqnarray}
b(a)&=&\Big(\lambda^2-\frac{\kappa^2\dot{\phi}^4}{4}\Big)^{1/4}a
=\Big(\lambda^2-\frac{\kappa^2C^4}{4a^{12}}\Big)^{1/4}a,\\
M(a)&=&\frac{\lambda+\frac{\kappa\dot{\phi}^2}{2}}{\Big(\lambda^2-\frac{\kappa^2\dot{\phi}^4}{4}\Big)^{1/4}}
=\frac{\lambda+\frac{\kappa C^2}{2a^6}}{\Big(\lambda^2-\frac{\kappa^2C^4}{4a^{12}}\Big)^{1/4}}.
\end{eqnarray}
Introducing the dimensionless variables $\bar{a}\equiv a/\sqrt{C}$, $\bar{b}\equiv b/\sqrt{C}$, $\bar\tau\equiv\tau/\sqrt{C}$ and $\bar\kappa\equiv\kappa/C$, the above equations can be simplified as
\begin{eqnarray}
\bar{b}(\bar{a})&=&\Big(\lambda^2-\frac{\bar\kappa^2}{4\bar{a}^{12}}\Big)^{1/4}\bar{a},\label{vilenba}\\
M(\bar{a})&=&\frac{\lambda+\frac{\bar\kappa}{2\bar{a}^6}}{\Big(\lambda^2-\frac{\bar\kappa^2}{4\bar{a}^{12}}\Big)^{1/4}}\label{vilenm}.
\end{eqnarray}
It can be seen that $\bar{a}$ is bounded from below:
\begin{equation}
\bar{a}(\bar\tau)\ge\Big(\frac{|\bar\kappa|}{2\lambda}\Big)^{1/6}\equiv\bar{a}_0,
\end{equation}
due to the presence of fourth roots. Assuming the asymptotically flat boundary condition \eqref{vilenboundarycondition}, the numerical solutions of $\bar{a}(\bar\tau)$ and $\phi(\bar\tau)$ as functions of $\bar\tau$ are shown in the dashed curves of Figure \ref{vilena} and Figure \ref{vilenphi}, respectively, for $\bar\kappa$ being positive (left) and negative (right). Note that we assume $\lambda=1$ (recall the effective cosmological constant is $(\lambda-1)/\kappa$) in these numerical calculations for the sake of comparison with the GR counterpart. The solid black curves in each figure represent Vilenkin's solution mentioned in subsection \ref{subsec:grvil}, i.e., the solutions in the framework of GR. In GR, the scale factor $\bar{a}$ vanishes at $\bar{\tau}_s$, a singular point that is highlighted by the vertical solid lines. At the singularity, the scalar field diverges logarithmically as shown in Eqs.~\eqref{grvilenkinasy}. However, in the EiBI theory with a positive $\bar{\kappa}$, the scale factor reaches its minimum $\bar{a}_0$ at a certain $\bar{\tau}$, which is highlighted by the vertical dashed line. The scalar field terminates at this point as well, with its derivative being a non-zero value. On the other hand, if $\bar\kappa<0$, the scale factor reaches $\bar{a}_0$ when $\bar\tau\rightarrow-\infty$. The scalar field diverges linearly in this limit. The detailed scrutiny on the asymptotic solutions near $\bar{a}_0$ will be presented in the following two subsubsections.

\subsubsection{Asymptotic behavior near $\bar{a}_0$: positive $\bar\kappa$}\label{subsubposik}
To investigate the asymptotic behavior of the solutions near $\bar{a}_0$, we consider the region where the Euclidean scale factor is very close to the minimum and expand it as follows:
\begin{equation}
\bar{a}(\bar\tau)=\bar{a}_0[1+\delta(\bar\tau)],
\end{equation}
where $0<\delta(\bar\tau)\ll 1$. If we assume $\bar\kappa>0$, Eqs.~\eqref{vilenba} and \eqref{vilenm} can be approximated as
\begin{equation}
\begin{split}
\bar{b}(\bar{\tau})\approx[12\lambda^2\delta(\bar\tau)]^{1/4}\bar{a}_0\equiv\Delta^{1/4}\bar{a}_0\,,
\qquad
M(\bar\tau)\approx\frac{2\lambda}{\Delta^{1/4}}\,,
\end{split}
\label{pkb}
\end{equation}
where $\Delta\equiv12\lambda^2\delta(\bar\tau)$. Furthermore, Eq.~\eqref{ee2} is approximated as
\begin{equation}
\dot{b}=\frac{d\bar{b}}{d\bar{a}}\frac{d\bar{a}}{d\bar\tau}\approx\Big(\frac{3\lambda^2}{\Delta^{3/4}}\Big)\frac{d\bar{a}}{d\bar\tau}\approx\frac{2\lambda}{\Delta^{1/4}}\sqrt{1+\frac{\bar{a}_0^2}{2\bar\kappa}}.
\end{equation}
Then we have
\begin{equation}
\begin{split}
\frac{d\Delta}{d\bar\tau}\propto\sqrt{\Delta}\,,
\qquad
\bar\tau-\bar\tau_0\propto\sqrt{\Delta}\,,
\end{split}
\end{equation}
where $\bar\tau_0$ satisfies $\bar{a}(\bar\tau_0)=\bar{a}_0$.

We therefore obtain the approximate solution of $\bar{a}(\bar\tau)$ near $\bar{a}_0$ as
\begin{equation}
\bar{a}(\bar\tau)\approx\bar{a}_0[1+\mathcal{A}(\bar\tau-\bar\tau_0)^2],
\label{vilenapk}
\end{equation}
where
\begin{equation}
\mathcal{A}=\frac{4}{3\bar{a}_0^2}\Big(1+\frac{\bar{a}_0^2}{2\bar\kappa}\Big)=\frac{2}{3\bar\kappa}+\frac{4}{3}\Big(\frac{2\lambda}{\bar\kappa}\Big)^{1/3}.\nonumber
\end{equation}
The behavior of the scalar field $\phi(\bar\tau)$ at this limit can also be obtained with Eq.~\eqref{phidot}:
\begin{equation}
\phi(\bar\tau)\approx\phi_0-\frac{1}{\bar{a}_0^3}(\bar\tau-\bar\tau_0)[1-\mathcal{A}(\bar\tau-\bar\tau_0)^2].
\label{phivilenpkkk}
\end{equation}
Note that at $\bar\tau_0$, the scale factor $\bar{a}$ approaches its minimum $\bar{a}_0$ with $d\bar{a}/d \bar\tau\rightarrow0$. The scalar field $\phi$, on the other hand, approaches its maximum $\phi_0$ with a non-vanishing derivative, i.e., $d\phi/d\bar\tau\rightarrow-1/\bar{a}_0^3$. It is worth to stress that even though the physical metric is regular when $\bar{a}=\bar{a}_0$, this point is actually a curvature singularity of the auxiliary metric because $R[q]\propto(\bar\tau-\bar\tau_0)^{-1}$, with $b$ vanishing as in Eq.~\eqref{pkb}.

\subsubsection{Asymptotic behavior near $\bar{a}_0$: negative $\bar\kappa$}
If $\bar\kappa<0$, Eqs.~\eqref{vilenba} and \eqref{vilenm} can be approximated as follows:
\begin{equation}
\begin{split}
\bar{b}(\bar{\tau})\approx\Delta^{1/4}\bar{a}_0\,,
\qquad
M(\bar\tau)\approx\frac{\Delta^{3/4}}{2\lambda}\,,
\end{split}
\end{equation}
near $\bar{a}_0$. Then Eq.~\eqref{ee2} is approximated as
\begin{equation}
\dot{b}=\frac{d\bar{b}}{d\bar{a}}\frac{d\bar{a}}{d\bar\tau}\approx\Big(\frac{3\lambda^2}{\Delta^{3/4}}\Big)\frac{d\bar{a}}{d\bar\tau}\approx\sqrt{\frac{\bar{a}_0^2}{6|\bar\kappa|}}\Delta^{1/4}.
\end{equation}
We then have
\begin{equation}
\begin{split}
\frac{d\Delta}{d\bar\tau}\approx\sqrt{\frac{8}{3|\bar\kappa|}}\Delta\,,
\qquad
\Delta\propto\textrm{exp}{\sqrt{\frac{8}{3|\bar\kappa|}}\bar\tau}\,.
\end{split}
\end{equation}
Therefore, the approximate solution of $\bar{a}(\bar\tau)$ near $\bar{a}_0$ reads
\begin{equation}
\bar{a}(\bar\tau)\approx\bar{a}_0\Big\{1+\mathcal{B}\Big[\textrm{exp}\Big(\sqrt{\frac{8}{3|\bar\kappa|}\bar\tau}\Big)\Big]\Big\},
\label{vilenanka}
\end{equation}
where $\mathcal{B}$ is a positive constant. One can see that the minimum $\bar{a}_0$ is reached when $\bar\tau\rightarrow-\infty$. On the other hand, the behavior of the scalar field $\phi(\bar\tau)$ is:
\begin{equation}
\phi(\bar\tau)\approx-\frac{\bar\tau}{\bar{a}_0^3},
\label{phinkappa}
\end{equation}
when $\bar\tau\rightarrow-\infty$. Note that even though the physical metric is regular, the scalar curvature of the auxiliary metric diverges in the sense that $R[q]\propto\Delta^{-3/2}$ in this limit.

\subsubsection{The on-shell action}
The on-shell action of the instanton solutions in the EiBI theory is given by Eq.~\eqref{bulkaction} when $N=1$. For the Vilenkin instanton in the EiBI theory, the boundary contribution $S_B$ is from a boundary corresponding to the singularity of the auxiliary metric. It can be written as
\begin{equation}
S_B=-\int_{\partial M}d^3x\sqrt{+h(q)}K(q)=\frac{6\pi^2b^2}{M}\frac{db}{d\tau}\Big|_{\partial M}.
\end{equation}
If $\bar\kappa>0$, the boundary contribution $S_B$ vanishes because $S_B\propto\sqrt{\Delta}$. If $\bar\kappa<0$, on the other hand, we have 
\begin{equation}
S_B\Big|_{\bar\kappa<0}=\sqrt{12\lambda}\pi^2C,
\end{equation}
which is independent of $\bar\kappa$. Note that the boundary in this case is located at $\bar\tau\rightarrow-\infty$.

\begin{figure}[t]
\centering
\graphicspath{{fig/}}
\includegraphics[scale=1.0]{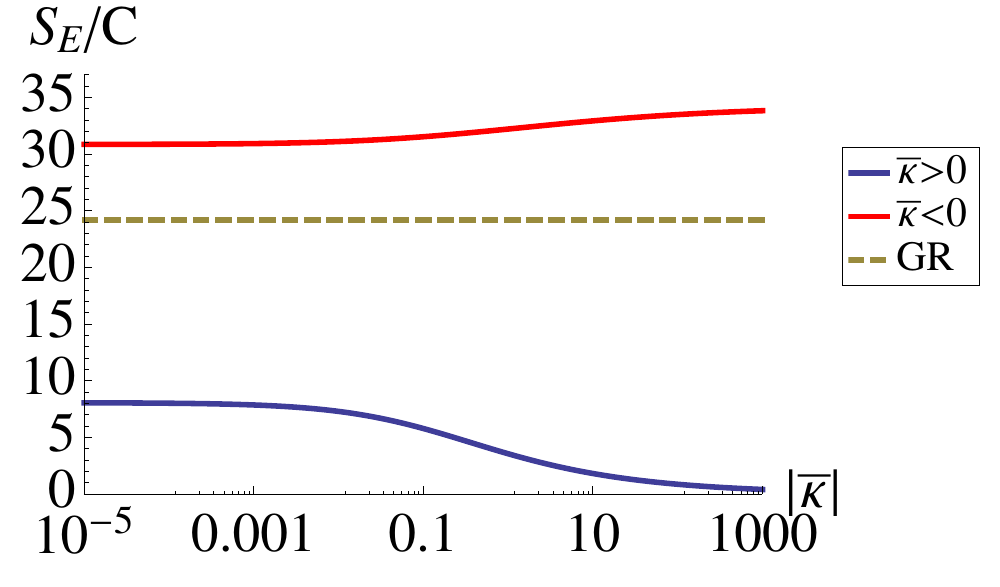}
\caption{The numerical results of the on-shell action as a function of $\bar\kappa$ for positive $\bar\kappa$ (blue) and for negative $\bar\kappa$ (red). The action value of the GR counterpart is also shown in the dashed line.}
\label{vilenactionnumerical}
\end{figure}

We consider the difference between the on-shell action of the instanton solution and that corresponding to a background flat spacetime:
\begin{equation}
\Delta S=S_E-S_{E(background)}.
\end{equation}
The solution of the background flat spacetime is 
\begin{equation}
\begin{split}
a(\tau)=b(\tau)=\tau\,,
\qquad
N=M=1\,,
\end{split}
\end{equation}
and the action of the background solution is zero. Note that the on-shell EiBI action of the background solution reduces to that in GR. Therefore, $\Delta S$ becomes
\begin{eqnarray}
\Delta S&=&\frac{2\pi^2}{\kappa}\int d\tau a^3\Big(\lambda+\frac{\kappa\dot{\phi}^2}{2}\Big)\Big(1-\frac{b^2}{a^2}\Big)+S_B\nonumber\\
&=&\frac{2\pi^2}{\kappa}\int_{a_0}^{\infty} da a^3\frac{[1-F(a)]\Big[F(a)+\frac{3}{4}\frac{\kappa^2C^4}{F(a)a^{12}}\Big]}{G(a)}+S_B,
\end{eqnarray}
where we define two new functions $F(a)$ and $G(a)$ as
\begin{equation}
\begin{split}
F(a)\equiv\sqrt{\lambda^2-\frac{\kappa^2C^4}{4a^{12}}}\,,
\qquad
G(a)\equiv\sqrt{1-\frac{a^2}{3\kappa}\Big[F(a)+\frac{F(a)^2}{2(\lambda+\frac{\kappa C^2}{2a^6})^2}-\frac{3}{2}\Big]}.
\end{split}
\end{equation}
After a change of variable $F(a)\equiv x$ and setting $\lambda=1$, we have
\begin{equation}
\Delta S=\frac{2\pi^2C}{\bar\kappa}\int_{0}^{1} \frac{dx}{6(1+x)}\Big[\frac{\bar\kappa^2}{4(1-x^2)}\Big]^{\frac{1}{3}}\frac{3-2x^2}{G_{\pm}(x)}+S_B,
\end{equation}
where the new function $G_{\pm}(x)$ is defined as
\begin{equation}
G_{\pm}(x)\equiv\sqrt{1-\frac{1}{3\bar\kappa}\Big[\frac{\bar\kappa^2}{4(1-x^2)}\Big]^{\frac{1}{6}}\Big[x+\frac{x^2}{2(1\pm\sqrt{1-x^2})^2}-\frac{3}{2}\Big]}.
\end{equation}
The plus (minus) sign corresponds to a positive (negative) $\bar\kappa$. Note that $\bar\kappa\equiv\kappa/C$ is dimensionless. The numerical results of the on-shell action as a function of $\bar\kappa$ are shown in Figure~\ref{vilenactionnumerical}. The blue and red curves are for positive and negative $\bar\kappa$, respectively. The value of the on-shell action of the singular Vilenkin instanton in GR is also shown by the dashed line. It can be seen that the Euclidean action has a well-defined value, hence one can conclude that the Minkowski space is still unstable even though the physical metric is regular. Moreover, an interesting observation is that the action does not converge to its GR counterpart as $|\bar\kappa|\rightarrow 0$. Since we find that the Minkowski space is at best metastable it is important to estimate the timescale of the instability, $t$. The decay rate per unit volume is given by $\gamma\sim 1/t^4\propto \mathrm{exp}(-\Delta S)$. \footnote{In fact, there should be a dimensionful prefactor $A$ with $[A]=1/[\mathrm{length}]^4$ in front of the exponential term. In the cases of false vacuum decay, the semiclassical estimation of $A$ was shown in Ref.~\cite{Callan:1977pt} in flat spacetime and in Ref.~\cite{Koehn:2015hga} when gravity is included. In \cite{Callan:1977pt} it was shown that $A\propto (\Delta S)^2$. Furthermore, if the action is sufficiently large in reduced Planck units ($\Delta S>1$), which is consistent with our estimation, the exponential term dominates over the prefactor. Given that $A\sim \mathcal{M}^4$, where $\mathcal{M}$ is the characteristic mass scale of the theory and we assume it to be $\mathcal{M}\sim 1/l_P$, we can estimate the decay rate by setting $A\sim \mathcal{O}(1)$ (in reduced Planck units). We will also assume this estimation is qualitatively valid in the EiBI gravity.} Given that our Universe is at least $10^{17}$ seconds old and the spacetime volume of our past light cone is roughly its fourth power, this implies that $\Delta S\gtrsim 560$ (in reduced Planck units) in order not to contradict observations. Assuming that $C\lesssim1$ (if the field excursion, which is proportional to $C$, is less than the Planck scale, it is reasonable to assume that $C\lesssim1$) \cite{Vilenkin:1998pp}, we find that the previous constraint is not satisfied for any value of $\kappa$ and consequently these instanton solutions are ruled-out observationally. Furthermore, from a theoretical perspective, these instantons are not totally satisfactory due to the presence of the singularity in the auxiliary metric. Therefore, from a conservative point of view, they should not be considered in the path integral of the EiBI theory. Note that there are some theoretical, but somewhat drastic, alternatives to explain this inconsistency and we will mention about them in the conclusion.

\section{Hawking-Turok singular instanton}\label{secHT}
The Hawking-Turok instanton in \cite{Hawking:1998bn} is based on the Hartle-Hawking no-boundary proposal to create an open universe from nothing \cite{Hartle:1983ai}. Unlike most of other instanton solutions in which a special shape of the scalar field potential is required, a generic potential is assumed in the Hawking-Turok proposal. However, the inevitable singularity for the curvature and the scalar field is the price that should be paid in this setup. As discussed in the introduction, even though the singularity is mild in the sense that the action can be finite as long as the potential is not too steep in $\phi$, the presence of the singularity has raised some criticisms and made this proposal controversial. 

\subsection{GR case}
\begin{figure}
\graphicspath{{fig/}}
\includegraphics[scale=0.9]{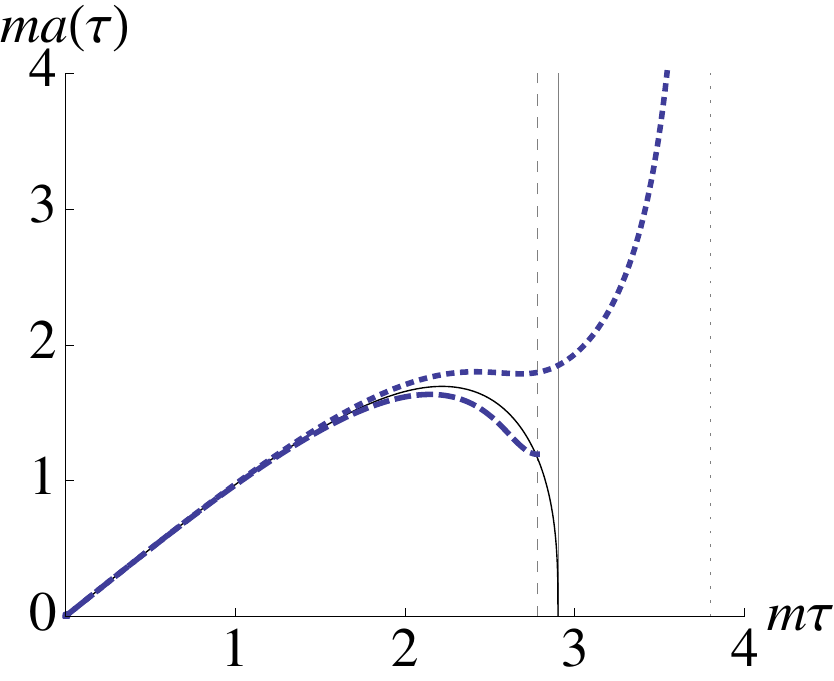}
\includegraphics[scale=0.9]{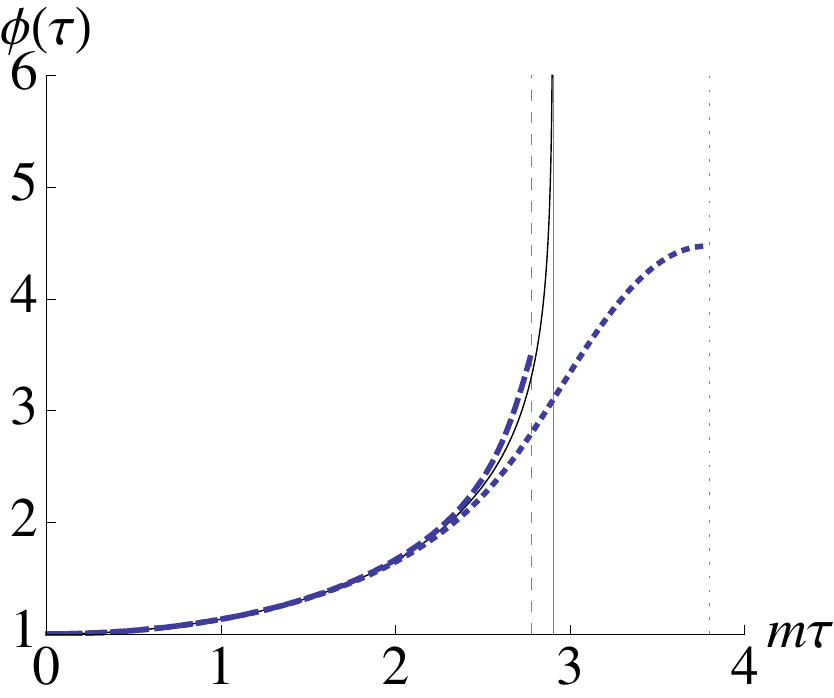}
\caption{The numerical solutions of $a(\tau)$ (left) and $\phi(\tau)$ (right) for the Hawking-Turok instanton in GR (black solid curves), in the EiBI theory with $\kappa=0.1/m^2$ (dashed curves) and $\kappa=-0.1/m^2$ (dotted curves). The scalar field potential is assumed to be a quadratic potential, $V=m^2\phi^2/2$. The initial condition for each cases is $\phi(0)=1$, $\dot{\phi}(0)\approx0$, $a(0)\approx0$ and $\dot{a}(0)\approx1$ at the north pole ($\tau=0$). The solid, dashed, and dotted vertical lines indicate the south poles $\tau_{sp}$ for GR, positive $\kappa$ and negative $\kappa$, respectively.}
\label{HTeibi}
\end{figure}

The initial condition for the Hawking-Turok instanton is $a(\tau)\approx\tau$ and $\dot{\phi}(\tau)\approx0$ at the north pole, i.e., when $\tau\rightarrow0$ \cite{Hawking:1998bn}. The initial condition of $\phi(0)$ can be set at will. The set of equations of motion in GR is
\begin{equation}
\begin{split}
\ddot{\phi}+3\frac{\dot{a}}{a}\dot{\phi}=\frac{\partial V}{\partial\phi}\,,
\qquad
\ddot{a}=-\frac{a}{3}(\dot{\phi}^2+V(\phi))\,.
\end{split}
\end{equation}
Note that the Euclidean evolution corresponds to the usual Lorentzian evolution in a potential $-V(\phi)$. The numerical solutions for a quadratic potential $V=m^2\phi^2/2$, where the mass $m$ is a constant, are shown in the black solid curves in Figure~\ref{HTeibi}. At the north pole, the solutions are regular. However, near the south pole at $\tau=\tau_{sp}$, the anti-damping term $\dot{\phi}\dot{a}/a$ dominates over the potential term. This implies that the behavior of the solutions near the south pole is similar to those of Vilenkin's solutions near the singularity. In other words, $a(\tau)$ vanishes, $\phi(\tau)$ diverges logarithmically at $\tau=\tau_{sp}$ and the Ricci scalar diverges. We expect this singular behavior can be modified in the EiBI theory and this is the main content of the next subsection.

\subsection{EiBI case}
In this subsection, we turn to the investigation of the Hawking-Turok instanton in the EiBI theory. If $\kappa>0$, the asymptotic behavior of the solutions near the south pole $\tau_{sp}$ can be estimated by assuming
\begin{equation}
\lambda+\kappa V(\phi)-\frac{\kappa\dot{\phi}^2}{2}=0.
\end{equation}
Differentiating this equation and substituting it into Eq.~\eqref{eibiphi} we obtain
\begin{equation}
\begin{split}
\ddot{\phi}=\frac{dV}{d\phi}\,,
\qquad
\frac{\dot{a}}{a}=0\,,
\end{split}
\end{equation}
at $\tau_{sp}$. As an example, consider a quadratic scalar field potential, $V=m^2\phi^2/2$. Then the asymptotic solution of $\phi$ can be described by
\begin{equation}
\phi(\tau)\Big|_{\tau\rightarrow\tau_{sp}}=\frac{1}{2\mathcal{C}}\textrm{e}^{m(\tau-\tau_{sp})}-\frac{\lambda\mathcal{C}}{\kappa m^2}\textrm{e}^{-m(\tau-\tau_{sp})}\approx\frac{1}{2\mathcal{C}}-\frac{\lambda\mathcal{C}}{\kappa m^2},
\end{equation}
where $\mathcal{C}$ is an integration constant. Therefore, the scalar field $\phi$ and its derivative $\dot{\phi}$ are non-zero and finite at the south pole. The physical scale factor $a$, on the other hand, approaches a constant with its derivative going to zero at $\tau_{sp}$. See the dashed curves in Figure~\ref{HTeibi}. We would like to emphasize that just like the case for the Vilenkin instanton in the EiBI theory, the scalar curvature of the physical metric is regular and in this sense the singularity of the Hawking-Turok instanton is removed. However, the scalar curvature of the auxiliary metric $R[q]$ still diverges at the south pole.

If $\kappa<0$, we find that the solutions behave quite differently from those mentioned previously. For a quadratic potential, we find that at the south pole, the auxiliary scale factor $b$ terminates at a finite but non-zero value. Furthermore, we also find that at $\tau_{sp}$,
\begin{equation}
\begin{split}
\lambda-|\kappa|V(\phi)-\frac{|\kappa|\dot{\phi}^2}{2}\rightarrow0\,,
\qquad
\lambda-|\kappa|V(\phi)+\frac{|\kappa|\dot{\phi}^2}{2}\rightarrow0\,.
\end{split}
\label{5.5}
\end{equation}
This implies that $\dot{\phi}\rightarrow0$ at the south pole. However, the ratio between the equations in \eqref{5.5} is a non-zero and finite constant. According to this numerical behavior, we obtain the asymptotic behavior of $a$ and $\phi$ near the south pole as follows:
\begin{equation}
\begin{split}
a(\tau)\propto(\tau_{sp}-\tau)^{-1}\,,
\qquad
\phi(\tau)\approx\phi_{sp}-\mathcal{D}(\tau_{sp}-\tau)^2\,,
\end{split}
\end{equation}
where $\mathcal{D}$ is a positive constant and $\phi_{sp}=\phi(\tau_{sp})$.
The numerical solutions of $a$ and $\phi$ are shown in the dotted curves of Figure~\ref{HTeibi}. One can see that $a$ diverges at the south pole and this divergence corresponds to a singular boundary of the physical metric, even though the scalar field is regular there. The scalar curvature of the auxiliary metric also diverges at the south pole.

\begin{figure}[t]
\centering
\graphicspath{{fig/}}
\includegraphics[scale=0.8]{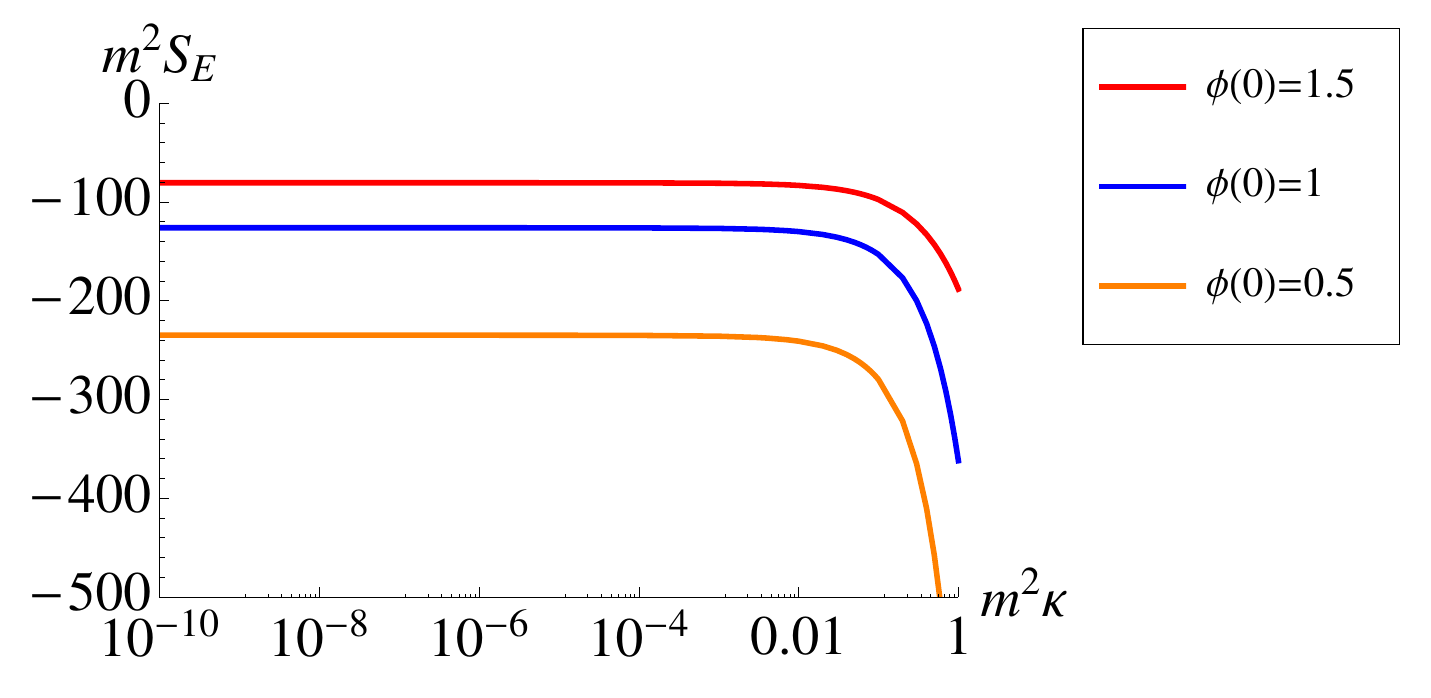}
\includegraphics[scale=0.8]{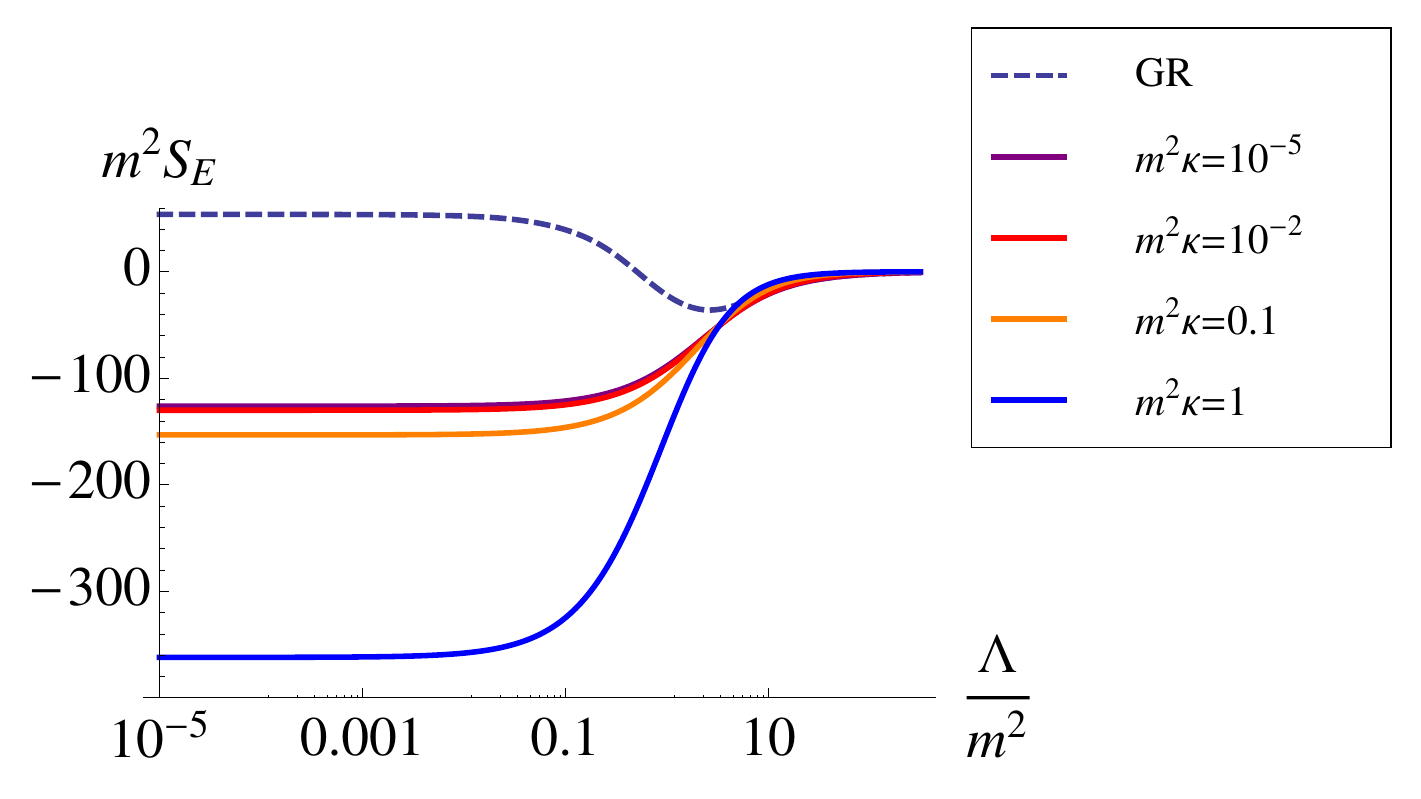}
\caption{In these two plots we show the on-shell action of the Hawking-Turok instanton in the EiBI theory for $\kappa>0$. On the top plot we assume a zero cosmological constant, i.e., $\lambda=1$ and compare the on-shell action as a function of $\kappa$ with different initial condition for the scalar field, $\phi(0)$. On the bottom plot, we fix the initial scalar field value to $\phi(0)=1$ and show the on-shell action as a function of the positive cosmological constant, for different values of $\kappa$. The on-shell action of the GR de-Sitter counterpart is also shown in the dashed curve.  Note that the boundary contribution $S_B$ is zero when $\kappa>0$.}
\label{HTeibiactionp}
\end{figure}

The on-shell action is directly related to the probability for a given initial state of the universe to tunnel to the solution described above. If $\kappa>0$, the boundary contribution $S_B$ is zero because the asymptotic behavior of the solutions are similar to those in the Vilenkin instanton in the EiBI theory, in which the boundary contribution is also zero. The on-shell action is then given by the bulk term and is shown in Figure~\ref{HTeibiactionp}. In the top plot, we assume a zero cosmological constant, i.e., $\lambda=1$ and compare the on-shell action as a function of $\kappa$ with different initial condition, $\phi(0)$. One can see that the solution with a smaller initial scalar field has a smaller on-shell action value and is exponentially preferred. In the bottom plot, we fix the initial value of the scalar field to $\phi(0)=1$ and assume a positive cosmological constant, i.e., $\lambda>1$. We compare the on-shell action value as a function of $\Lambda$ with different $\kappa$. The Hawking-Turok GR counterpart is also shown in this plot (see the dashed curve). One can see that all the action values approach zero once the cosmological constant is getting large. Most importantly, different from the GR curve in which there is a local minimum for a certain $\Lambda\approx m^2$, the EiBI theory seems to prefer a smaller cosmological constant. Furthermore, it can be seen that for a small enough value of $\kappa$, the curves will converge toward a specific curve (see the region around the purple and red curves in the middle). This specific curve is not the one calculated in GR. Therefore, the Euclidean action does not converge to its GR counterpart when $\kappa\rightarrow 0$. This phenomenon is rather similar to what we have found in the Vilenkin instanton within the EiBI setup (see Figure~\ref{vilenactionnumerical}).

In the case that $\kappa<0$, we found that the solutions possess a divergent action value due to the presence of the singularities of the physical metric and the auxiliary metric. Because of this fact, we dismiss the $\kappa<0$ case as being unphysical.

In summary, we found that for the Hawking-Turok instanton on a quadratic potential in the EiBI theory, the singularity of the physical metric can be removed when $\kappa>0$ in the sense that the scalar curvature constructed from the physical metric, the scale factor $a$ and the scalar field $\phi$ are regular at the south pole. However, the scalar curvature constructed from the auxiliary metric, i.e., $R[q]$, diverges at this pole. Furthermore, we found that the on-shell action for $\kappa>0$ is given by the bulk term and is finite. However, the action value does not converge to the GR value in the appropriate limit, i.e., $\kappa\rightarrow 0$. On the other hand, at the south pole there is a singularity of the physical metric when $\kappa<0$. The scalar curvature constructed from both metrics and the physical scale factor $a$ are singular at this pole. The action value is not well-defined for this case as well. Therefore, if our universe is created from a Hawking-Turok instanton, then it should have a positive $\kappa$. Recall that in Ref.~\cite{Avelino:2012ge} it was found that there may be unwanted instabilities resulting from the imaginary effective sound speed in the EiBI theory with a negative $\kappa$.

Before closing this section, we would like to stress that it is still unclear whether the Hawking-Turok instanton solutions with a positive $\kappa$ should be excluded from the path integral (the solutions with negative $\kappa$ should be excluded because the action is not well-defined). From a theoretical point of view, one may suggest that these solutions should be excluded because they are singular in the auxiliary metric. In addition, due to the similar auxiliary singularities shared by the Vilenkin instantons, the validity of the Hawking-Turok instantons is challenged based on the observational inconsistency predicted by the Vilenkin instantons in the EiBI theory. We need to keep in mind that these arguments are not formal proofs to exclude the Hawking-Turok instantons and a conclusive answer is still needed.

\section{Conclusion}\label{sec:conclusion}
In this paper, we investigated three types of $O(4)$-symmetric instantons in the EiBI gravity theory.

For the Hawking-Moss instanton we found that the tunneling rate reduces to that of the Hawking-Moss instanton in GR for all $\kappa$, while the action values reduce to their GR counterpart only in the limit when the EiBI theory is expected to approach GR, i.e., $\kappa\rightarrow 0$. Interestingly, we also found that the well-known Bekenstein-Hawking entropy formulae is not satisfied (there is an additional term proportional to $\kappa$ in Eq.~\eqref{hmentropy}) if one uses the physical metric to define the area of the cosmological horizon, while it is satisfied if one uses the area of the cosmological horizon of the auxiliary metric (recall we define the physical metric as the one that is minimally coupled with the scalar field). 

We then studied two types of singular instantons in the EiBI theory: the Vilenkin instanton and the Hawking-Turok instanton.

The Vilenkin instanton setup is defined in the Minkowski background with a free scalar field. As we expected, the quantum mechanical extension of the EiBI theory via instanton dynamics retains its property of resolving the singularity in the sense that the curvature of the physical metric is regular (see Figures~\ref{vilena} and \ref{vilenphi}) but the curvature of the auxiliary metric is still singular. Since the instanton tunneling description gives a well-defined probability (see Figure~\ref{vilenactionnumerical}), our result implies that the Minkowski space is still unstable in the EiBI theory. In this theory, the problem of whether one should accept these instantons or not has to be treated more carefully than in GR because the theory partially \textit{resolves} the singularity of the instanton.

Assuming that the integration constant $C$ defined in Eq.~\eqref{phidot} is smaller than one (in Planck units), we find that for any value of $\kappa$ the timescale of the instability is unacceptably small compared with the age of our universe. We conclude that the Vilenkin instanton should be excluded from the Euclidean path integral of the EiBI theory because it grossly contradicts observations. In addition, one could also take a conservative view and argue that from a theoretical point of view, the presence of the curvature singularity in the auxiliary metric already constitutes a sufficient condition to exclude this instanton from the path integral. 
There are possible ways to avoid the above conclusions but they are much more drastic in our opinion. For example, if one wishes to accept the Vilenkin instanton as legitimate then one is forced to either conclude that the EiBI theory is ruled-out (recall that in GR, the Vilenkin instanton also implies the observationally unacceptable fast decay of the Minkowski spacetime) or that the Euclidean path integral approach is incomplete or should be abandoned. Finally, another way out is given that the Minkowski space is unstable in the EiBI theory one could imagine that there exists some currently unknown physical mechanism that prevents the decay.

\begin{figure}[t]
\centering
\graphicspath{{fig/}}
\includegraphics[scale=0.8]{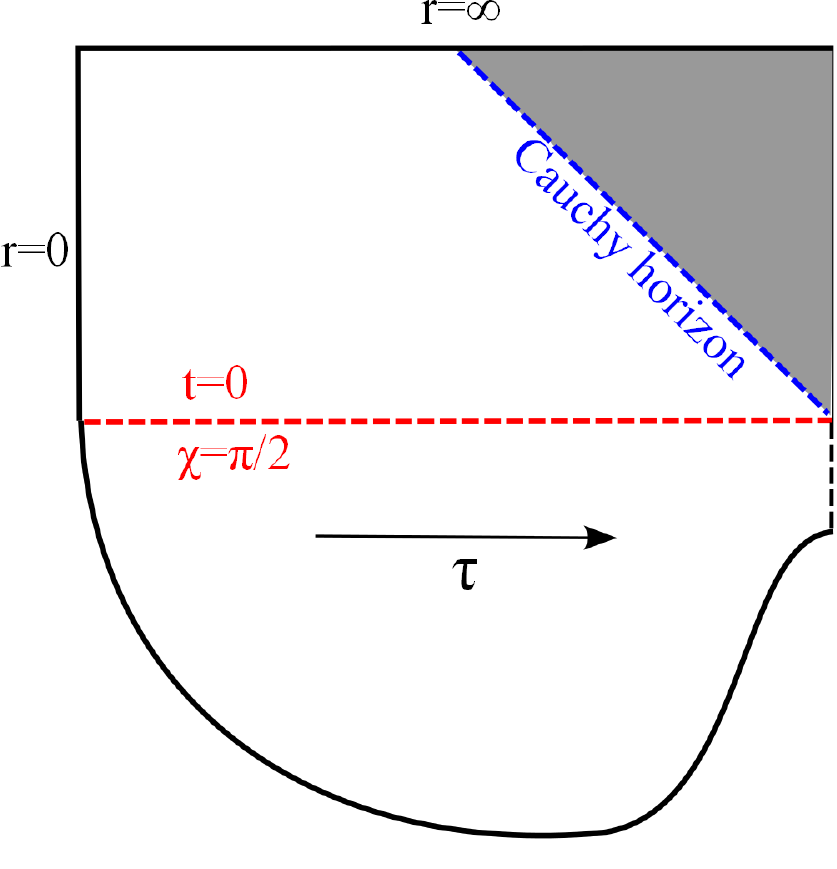}
\caption{In this figure we show the analytic continuation of the physical metric of the Hawking-Turok instanton in the EiBI theory with positive $\kappa$. The lower half part is the Euclidean section and the upper half part is the Lorentzian section. We analytically continue from the $\chi = \pi/2$ hypersurface of the Euclidean manifold to the Lorentzian manifold. Since the right end of the Euclidean manifold is non-compact (while the Hawking-Turok instanton in GR would be compact), after analytic continuation, there appears a Cauchy horizon. However, the physical metric of the Euclidean and Lorentzian sections are both regular; this is the different point compared to GR.}
\label{penrose}
\end{figure}

We also investigated the Hawking-Turok instanton in the EiBI theory (see Figure~\ref{HTeibi}). For the positive $\kappa$ case, the singularity is also partially resolved in the EiBI theory and the probability is well-defined. The analytic continuation of the physical metric from the Euclidean section to the Lorentzian section is exhibited in Figure~\ref{penrose}. We found that the right end of the Euclidean manifold is non-compact and there appears a Cauchy horizon on the Lorentzian manifold after the analytic continuation. On the other hand, for the negative $\kappa$ case, the singularity is not resolved and the Euclidean action is divergent. We conclude that if our universe has been created via a Hawking-Turok instanton, then it should have positive $\kappa$. We would like to repeat that, similar to the Vilenkin instanton in the EiBI theory, the legitimacy of these instantons can be questioned due to the presence of the curvature singularity of the auxiliary metric.

One interesting finding is that the Euclidean actions of the instantons do not exactly approach the GR value in the appropriate limit. In the classical sense, if there is no scalar field, then whatever the value of $\kappa$ is, the solution should be the same as that of GR. However, as can be seen in the Hawking-Moss instanton in the EiBI theory, the action value depends on $\kappa$ and it only approaches the value in GR in the $\kappa \rightarrow 0$ limit. Moreover, once we consider singular instantons, then (even though the singularity is partially resolved via the EiBI theory) the probability does not approach to the GR value even as $\kappa$ goes to zero. This is not only true for the Vilenkin instanton, but also for the Hawking-Turok instanton. It is not clear to us whether this shows any inconsistency of singular instantons or not.

Finally, we suggest a possible extension of our conclusion to the Big Bang singularity issue. If we cannot accept singular instantons since there remains a singularity in the auxiliary metric, then we also have to be very careful about the avoidance of the initial singularity in the EiBI theory even though its physical metric is regular. To be sure, the resolution of the initial singularity problem via the EiBI theory should be examined not only at the classical level but also at the quantum level. We leave this issue for our future investigations.

\acknowledgments

CYC would like to thank Pao-Wen Yang for his help on this work. FA and DY are supported by the National Taiwan University (NTU) under Project No. 103R4000 and by the NTU Leung Center for Cosmology and Particle Astrophysics (LeCosPA) under Project No. FI121. CYC and PC are supported by Taiwan National Science Council under Project No. NSC 97-2112-M-002-026-MY3 and by Taiwan National Center for Theoretical Sciences (NCTS). PC is in addition supported by US Department of Energy under Contract No. DE-AC03-76SF00515.

\end{document}